\documentclass[sigconf]{acmart}

\AtBeginDocument{%
  }

\setcopyright{none}
\copyrightyear{}
\acmYear{2025}
\acmDOI{}
\acmConference[CHI '25 T4T Workshop]{CHI 2025 Workshop on Tools for Thought: Research and Design for Understanding, Protecting, and Augmenting Human Cognition with Generative AI}{April 26, 2025}{Yokohama, Japan}
\acmBooktitle{CHI~'25 T4T: Workshop on Tools for Thought---Research and Design for Understanding, Protecting, and Augmenting Human Cognition with Generative AI, April 26, 2025, Yokohama, Japan}
\acmISBN{}

\usepackage{soul}

\begin{document}

\title{Augmenting Human Cognition With Generative AI: Lessons From AI-Assisted Decision-Making}

\author{Zelun Tony Zhang}
\email{zhang@fortiss.org}
\orcid{0000-0002-4544-7389}
\affiliation{%
  \institution{fortiss GmbH, Research Institute of the Free State of Bavaria}
  \city{Munich}
  \country{Germany}
}
\affiliation{%
  \institution{LMU Munich}
  \city{Munich}
  \country{Germany}
}

\author{Leon Reicherts}
\email{leon.reicherts@microsoft.com}
\orcid{0000-0002-7338-5242}
\affiliation{%
 \institution{Microsoft Research}
 \city{Cambridge}
 \country{United Kingdom}
}


\begin{abstract}
  How can we use generative AI to design tools that augment rather than replace human cognition? In this position paper, we review our own research on AI-assisted decision-making for lessons to learn. We observe that in both AI-assisted decision-making and generative AI, a popular approach is to suggest AI-generated end-to-end solutions to users, which users can then accept, reject, or edit. Alternatively, AI tools could offer more incremental support to help users solve tasks themselves, which we call process-oriented support. We describe findings on the challenges of end-to-end solutions, and how process-oriented support can address them. We also discuss the applicability of these findings to generative AI based on a recent study in which we compared both approaches to assist users in a complex decision-making task with LLMs.
\end{abstract}

\ccsdesc[500]{Human-centered computing~Interaction paradigms}
\ccsdesc[500]{Human-centered computing~HCI theory, concepts and models}
\ccsdesc[300]{Information systems~Decision support systems}

\keywords{Generative AI, tools for thought, AI-assisted decision-making, process-oriented support, forward reasoning}


\maketitle

\section{Introduction}
In recent years, the perspective that AI should not replace, but augment and enhance human cognition has become increasingly popular~\cite{carter_using_2017,akata_research_2020,shneiderman_human-centered_2020}. Now, with the rapid progress in generative AI (GenAI), this perspective becomes even more relevant, as AI can generate satisfactory, or sometimes even impressive solutions to more and more tasks with little human effort. But how can we leverage these capabilities without sidelining human reasoning? Or more ambitiously, how can we combine artificial with human intelligence to produce better results than either could on its own? These questions have been extensively researched in the field of \textit{AI-assisted decision-making} as one of the primary applications of human-AI collaboration prior to the popularization of GenAI. Therefore, to understand how to design GenAI systems to protect and augment human thinking, it seems natural to turn to AI-assisted decision-making for lessons to learn. In this position paper, we review our own work in AI-assisted decision-making for how it can inform the design of GenAI tools for thought. We start by outlining several challenges of using AI to propose end-to-end solutions (\autoref{sec:end-to-end}), continue by presenting alternative approaches and how they can overcome these challenges (\autoref{sec:process}), and end by discussing the applicability of these findings to GenAI, based on our recent study on supporting complex decisions with LLMs (\autoref{sec:GenAI}).

\section{Challenges of End-to-End Solutions}
\label{sec:end-to-end}
There is a tendency in AI-assisted decision-making to design AI in a recommendation-centric way: The AI provides an end-to-end recommendation, i.e., a recommendation for the final decision straight from the input data, without involving the user. The user can then decide whether to accept or overrule this recommendation. A similar tendency can be observed in GenAI as well, where the AI often assists by generating an end-to-end solution to the problem, and the user is only involved afterwards as an editor. In our research, we identified several challenges that come with this approach. 

\textbf{Increasing overreliance over time.} In~\cite{zhang_is_2023}, we measured how participants' overreliance developed over repeated interactions with a recommendation-centric AI, and found that overreliance increased over time. This was to be expected for long, tiring study sessions that are routinely used in AI-assisted decision-making research. However, overreliance even increased when participants only had short interactions with the AI, which were spread over a longer period. The results indicate that providing end-to-end recommendations makes it difficult for users to remain meaningfully engaged with the decision-making task, similar to the difficulty of supervisory control in automation~\cite{bainbridge_ironies_1983}. 

\textbf{Overreliance in difficult tasks.} We conducted two studies using different methods and decision tasks to study how users' reliance on recommendation-centric AI depends on decision difficulty~\cite{zhang_you_2024,zhang_measuring_2024}, as the vision is often for AI to augment human reasoning in \textit{difficult} tasks. The results of both studies were consistent in that participants could often rely appropriately on AI recommendations in easy decisions, but became much more over-reliant in difficult decisions. The problem is that for decisions where users are highly uncertain about their answer, recommendation-centric support does little to improve their understanding of the decision. Consequently, users cannot meaningfully contribute and have little choice but to rely on the AI recommendation.

\textbf{Helpful when correct, but a burden when wrong.} A common assertion in the discourse about using AI in high-stakes domains is that AI models are black boxes that need to be opened. However, when we interviewed professional pilots about their thoughts on deploying AI into cockpits~\cite{zhang_pilot_2021}, transparency was not their main concern. Instead, they were much more concerned that AI might not be intelligent enough to respond appropriately to complex situations and become a burden when workload is already high---a phenomenon known as \textit{clumsy automation} in automation research~\cite{cook_cognitive_1991}. This finding points to a seemingly trivial, but often neglected requirement for the design of AI systems: that imperfect AI outputs should have minimal negative consequences for users, which \citet{gu_ai-resilient_2024} call \textit{AI-resilient}. In a follow-up study~\cite{zhang_resilience_2023}, we found why recommendation-centric support tends to violate this requirement. Pilots stated that they would not blindly trust AI, but they also rejected the idea that they would have to review every AI recommendation for its correctness. This seeming contradiction highlights that instead of augmenting users' reasoning, end-to-end recommendations burden users with an effortful review task that is not integrated into their existing reasoning processes---and one that becomes more effortful the more complex the decision is.

\section{Benefits of Process-Oriented Support}
\label{sec:process}
\begin{figure*}
    \centering
    \includegraphics[width=0.9\linewidth]{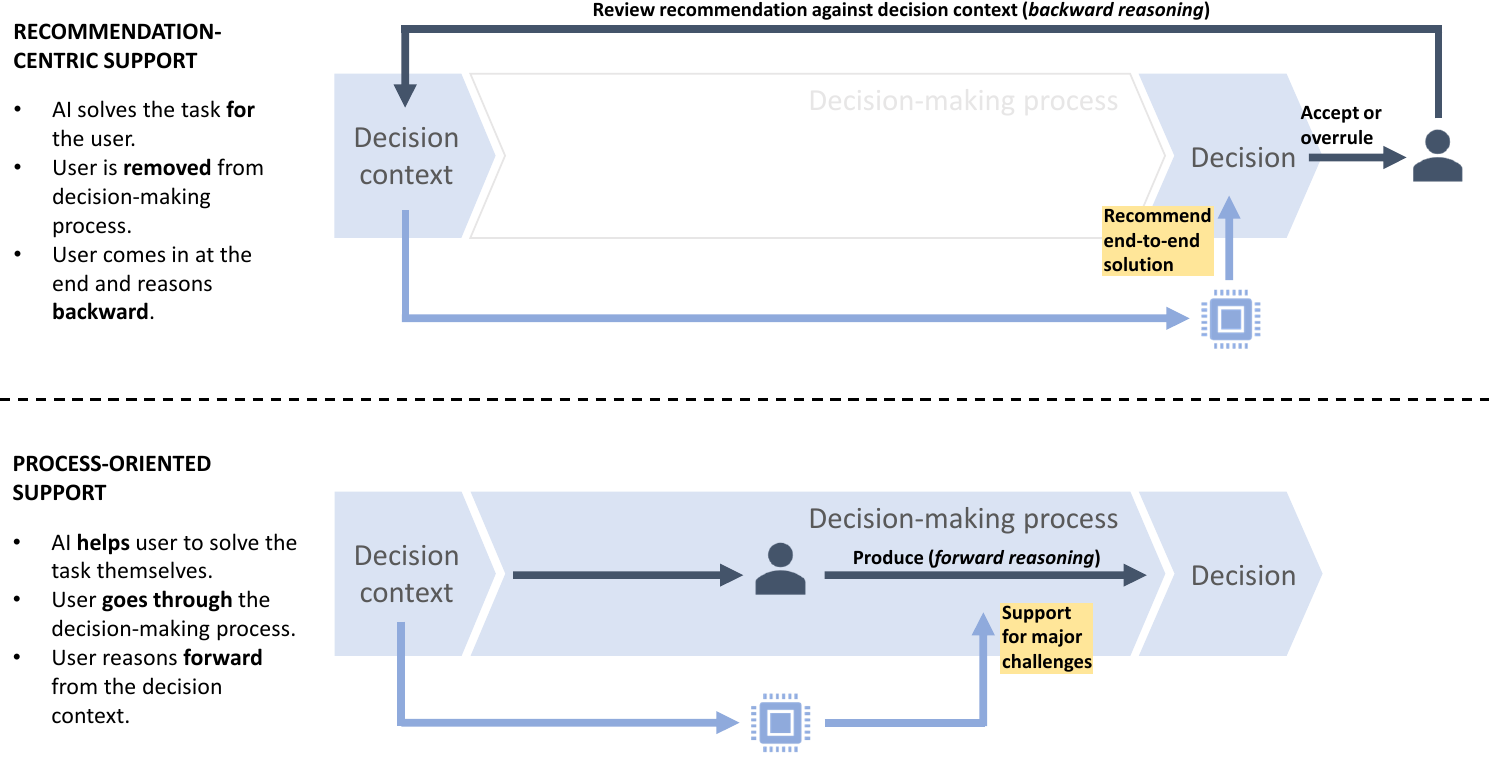}
    \caption{Recommendation-centric vs. process-oriented support. Adapted from \cite{zhang_beyond_2024}.}
    \Description{}
    \label{fig:process_support}
\end{figure*}
Providing end-to-end solutions is often the most obvious approach, but it is not the only one. Rather than solving tasks \textit{for} users, we can also aim to help users solve the task \textit{themselves}~\cite{dix_designing_2007} by identifying and focusing on their primary challenges when solving a task. Rather than end-to-end solutions, the AI would provide more incremental support targeted at these primary challenges. The goal would be to help users reason \textit{forward} through their reasoning processes, rather than pushing them to reason \textit{backward} from an end-to-end solution. We call this alternative paradigm \textit{process-oriented support} (\autoref{fig:process_support}).

We empirically compared recommendation-centric to process-oriented support in the use case of diversions in commercial aviation~\cite{zhang_beyond_2024}. When making diversion decisions, pilots' primary challenge is to gather and integrate the required information from multiple sources. Our process-oriented support concept addressed this challenge by pulling relevant information together in a continuously updated table view. AI continuously evaluated the information and highlighted possible limitations at the surrounding airports within the table. We found that this continuous provision of local hints was better accepted by pilots, led to less overreliance, and resulted in similar decision times as recommendation-centric support. Interestingly, only the combination of recommendations and continuous provision of local hints led to faster decisions, as in that case, recommendations were better integrated into pilots' reasoning, which removed the need for the time-consuming review of the recommendation. This also shows that process-oriented support does not preclude using AI to present recommendations or complete solutions. However, process-oriented support emphasizes that \textit{if} recommendations are given, they have to be embedded into the process such that users are still encouraged to reason forward---either through recommendations that are easy to verify~\cite{zhang_measuring_2024,zhang_you_2024}, or through additional incremental support leading up to the final solution~\cite{zhang_beyond_2024}.

Process-oriented support can take very different forms depending on the task. In another study, we investigated stock investment decision-making ~\cite{reicherts_extending_2022}. One of the key challenges of this task is to stay consistent with one's strategy and not to act too emotionally in reaction to market events. To address this challenge, we designed a chatbot with the aim to scaffold the decision-making process of experienced stock investors by asking reflective questions. We found that these probing questions from the chatbot got investors to consider aspects of their decision which they previously had not adequately taken into account, giving them the opportunity to reflect on certain aspects of an intended investment before proceeding with it. Other examples of process-oriented decision support are often found in healthcare, e.g. \citet{lindvall_rapid_2021} (key challenge addressed with AI: finding small structures in huge images) or \citet{zhang_rethinking_2024} (key challenge: uncertainty about patients' future state).

Overall, these studies demonstrate that by targeting AI support at the core challenges in users' existing working processes, we can design AI tools that engage and integrate much more with users' own reasoning than end-to-end solutions. As a result, process-oriented support---if properly designed---can address all of the challenges of end-to-end solutions laid out in \autoref{sec:end-to-end}. For one, as users' contributions remain key to producing a result, they are less likely to disengage from the task and become increasingly over-reliant on AI over time. Process-oriented AI tools further aim to enhance users' own understanding of the problem, which in turn should also help with making sense of and contextualizing AI support even in difficult decisions, leading to more appropriate reliance, as we could observe in our diversion use case~\cite{zhang_beyond_2024}. Lastly, since process-oriented support is integrated into users' reasoning, reviewing AI outputs becomes less effortful, as exemplified by the faster decisions when recommendations were combined with continuous provision of local hints~\cite{zhang_beyond_2024}. This reduces the burden when AI makes mistakes, provided subpar AI outputs can be easily dismissed or corrected.

\section{Applicability to Generative AI}
\label{sec:GenAI}
As the findings above are not unique to specific AI technologies, they likely also apply to many of the tasks GenAI tools are designed and used for. In fact, the challenges described in \autoref{sec:end-to-end} may become even more pronounced as GenAI becomes increasingly capable of solving more and more complex tasks end-to-end. At the same time, GenAI offers novel possibilities to design AI tools in a process-oriented manner. For instance, while the investment chatbot mentioned in \autoref{sec:process} was effective at triggering reflections in investors, its rule-based question prompts were sometimes not sufficiently relevant for what they were thinking or intending to do at the given moment~\cite{reicherts_extending_2022}. Through their ability to process unstructured human reasoning, LLMs may be able to address this challenge and enable the design of AI tools that are more tightly integrated with users' thought processes.

We explored how lessons from AI-assisted decision-making with ``traditional'' AI apply to GenAI by comparing an end-to-end approach with a more process-oriented approach for a GenAI tool for a complex decision-making task~\cite{reicherts_ai_2025}. Participants had to invest in a selection of ETFs (exchange-traded funds) to match a given investor profile. They interacted with two types of GenAI support: \textit{RecommendAI} directly recommended a set of ETFs to invest in, along with a short explanation for each recommendation. \textit{ExtendAI} first asked participants to describe their own investment plan and their rationale for the intended trades. The AI would then extend the rationale by embedding feedback into it, with the aim of highlighting additional factors to consider and uncovering blind spots.

Our findings show that \textit{ExtendAI} is a promising way to conceptualize human-GenAI interactions in complex (decision-making) tasks. With \textit{ExtendAI}, participants achieved slightly better diversified portfolios with fewer trades than with \textit{RecommendAI}. This indicates that \textit{ExtendAI} helped participants better understand how to improve their plan in a targeted way (forward reasoning), while participants tended to review \textit{RecommendAI}'s recommendations individually with a less holistic understanding of their impact on their portfolio (backward reasoning). \textit{ExtendAI} was further perceived as better integrated with participants' reasoning. As a result, while with \textit{RecommendAI}, participants faced the question of whether to trust the recommendations or not, this was less of a concern with \textit{ExtendAI}. Participants could more easily make sense of its feedback as it was directly tied to their own thoughts. Lastly, with \textit{ExtendAI}, participants appeared to perceive more ownership of their decisions, while with \textit{RecommendAI}, they tended to blame the AI for unsatisfactory outcomes.

However, we also found \textit{RecommendAI} to have its advantages. In particular, it was perceived as more actionable and insightful, and to give more fresh ideas as it was not directly tied to participants' own thoughts. Overall, preference between the two types of AI support was evenly split among participants, highlighting trade-offs along various dimensions when moving along the spectrum between end-to-end and process-oriented tools for thought. Some of the dimensions which we found in our research are:
\begin{itemize}
    \item \textbf{Timing of the support}: AI tools can provide assistance earlier or later in the user's reasoning process. Support in the early stages (e.g., via recommendations) can help the user get started and explore fresh perspectives, but may create a strong anchor that inhibits independent human reasoning. Support at later stages (such as with \textit{ExtendAI}) can be embedded into the user's own reasoning, but when it comes ``too late'', it can be difficult for the user to accommodate suggestions from the AI that do not align with their own thinking/plans, as they have already ``settled'' on what they are going to do~\cite{reicherts_ai_2025}. A good sweet spot seems to be support while the user's thoughts are ``in the making,'' for example when the user has a rough idea of what to do but is still in the process of identifying the pros and cons for doing so.
    \item \textbf{Type of support}: AI tools can offer various types of support, from \textit{providing}, to \textit{curating}, to \textit{interpreting} data. The type of support needs to be chosen based on the application context (e.g., type of task, user expertise, AI performance). While AI developers are often biased toward data interpretation, this form of support requires especially careful consideration, as it can easily diminish user autonomy when not properly embedded into the user's thinking process~\cite{zhang_beyond_2024,reicherts_ai_2025,zhang_rethinking_2024,yildirim_multimodal_2024}. In our diversion use case~\cite{zhang_beyond_2024}, we found that a combination of support types---continuous provision of local hints (curation) and recommendations (interpretation)---can be effective.
    \item \textbf{Degree of externalization required from the user}: To integrate AI support with the user's reasoning, some degree of cognitive externalization is often required from the user, such as with \textit{ExtendAI}, where users had to describe their rationale first before the AI can then extend it. In that case, the externalization itself was also helpful to some participants to better think through their decisions (similar to so-called ``rubber-duck debugging'' and the self-explanation effect~\cite{chi_eliciting_1994}). However, externalizing one's thoughts can also be challenging and effortful. It is thus important to consider if the amount of effort required to externalize one's thinking, and the form in which this is done, is adequate with respect to what users might gain from it. For example, our participants suggested providing a direct manipulation interface to specify intended trades and relying only on natural language input to describe the rationale for each trade~\cite{reicherts_ai_2025}.
    \item \textbf{Degree of friction introduced into the process}: One possible goal of AI tools is to help the user correct their own mistakes or notice their own oversights. To that end, it may be necessary to introduce a certain degree of friction into the user's reasoning process, as with \textit{ExtendAI}~\cite{reicherts_ai_2025} or the probing questions from the chatbot in our stock investment study~\cite{reicherts_extending_2022}. However, it is important to find the right balance that helps the user to reflect on their own thinking without becoming too annoying or disruptive to the user.
\end{itemize}

All in all, the results of our ETF investment study are in line with findings from our previous work---that process-oriented support tools can integrate more deeply with human reasoning and even lead to better task outcomes than the end-to-end approach---while also surfacing nuanced considerations in the design of such tools. The study confirms the applicability of these findings to GenAI, at least within the context of decision-making tasks. But even moving beyond to other knowledge work tasks for which GenAI is commonly used, we are confident that the strategy of process-orientation can serve as a valuable starting point in cases where protecting and augmenting human cognition is desired. For instance, results similar to ours have been found in tasks such as ideation~\cite{qin_timing_2025}, coding~\cite{vaithilingam_expectation_2022}, or writing~\cite{li_value_2024}.

\section{Conclusion}
In summary, based on our previous work on AI-assisted decision-making, we suggest that to preserve and enhance human reasoning in GenAI applications, the common approach of solving complex reasoning tasks end-to-end may often be inappropriate. In many applications, it may be easier for users to engage with the task and contribute their own reasoning with incremental, process-oriented support that facilitates forward reasoning, which might even lead to better outcomes than backward reasoning from end-to-end solutions. In a first exploratory study, we investigated both approaches applied to LLMs in the context of complex investment decision-making, confirming the applicability of these findings to GenAI, while also highlighting nuances on how to integrate process-oriented GenAI tools with human reasoning. Going forward, we suggest that process-oriented support can be an effective framework to inform the design of both ``traditional'' AI-assisted decision-making tools but also GenAI-based tools for thought.


\bibliographystyle{ACM-Reference-Format}
\bibliography{references}


\end{document}